\documentclass[showpacs,aps,prb,twocolumn]{revtex4}

\usepackage{graphicx}
\usepackage{dcolumn}
\usepackage{colordvi}
\usepackage{subfigure}

\def\E{{\tilde{E}}}
\def\P{{\tilde{P}}}

\begin{document}


\title{
Theoretical study of ferroelectric potassium nitrate
}

\author{
Oswaldo Di\'eguez\cite{presentaddress}
and David Vanderbilt
}

\affiliation{
Department of Physics and Astronomy, Rutgers University,
Piscataway, New Jersey 08854-8019, USA
}

\date{\today}


\begin{abstract}
We present a detailed study of the structural behavior and
polarization reversal mechanism in phase III of KNO$_3$, an unusual
ferroelectric material in which the nitrate groups rotate during
polarization reversal.  This material was one of several studied in
a previous work
[O.~Di\'eguez and D.~Vanderbilt, Phys.\ Rev.\ Lett.\ {\bf 96}, 056401
(2006)]
where methods were described for computing curves of energy
versus electric polarization.
In the present work
we extend and systematize the previous first-principles calculations
on KNO$_3$, and analyze in detail a two-parameter model in which the
energy of the system is written as a low-order expansion in the
polarization and the nitrate group orientation.  We confirm that
this model reproduces the first-principles results for KNO$_3$ very
well and construct its parameter-space phase diagram, describing
regions where unusual triple-well potentials appear.  We also
present first-principles calculations of KNO$_3$ under pressure,
finding that its energy-versus-polarization curves change character
by developing a first-derivative discontinuity at zero polarization.
\end{abstract}

\vskip 2mm


\maketitle

\marginparwidth 2.7in
\marginparsep 0.5in

\def\odm#1{\marginpar{\Blue{\small OD: #1}}}
\def\dvm#1{\marginpar{\Red{\small DV: #1}}}

\columnseprule 0pt


\section{Introduction}

Potassium nitrate (KNO$_3$), also known as saltpetre, has long been used
as an ingredient in explosives and propellants, including `black powder'
and other early forms of gunpowder.
At room temperature and pressure, KNO$_3$ crystallizes in an aragonite
({\it Pnma}) phase with four formula units.
This phase is usually referred to as phase II, and its domain of existence
is illustrated in the phase diagram of Fig.~1 in Ref.~\onlinecite{Scott1987PRB}.
When this material is heated at atmospheric pressure, a transition to phase I
occurs.
The structure of phase I seems to have been the subject of a controversy for
years, with the current status being not much changed since the discussion
given twenty years ago by Scott {\em et al.}\ in
Ref.~\onlinecite{Scott1987PRB}.
The proposed structure has global $C_3$ symmetry but local $D_6$ symmetry.
Upon cooling at atmospheric pressure, phase I does not transform back to phase
II; instead there is a narrow window of temperatures (aproximately from
113$^\circ$C to 120$^\circ$C) in which a ferroelectric phase, known as phase
III, occurs.
It is possible to broaden the range of temperatures in which phase III exists
by applying hydrostatic pressure to the bulk system, or by growing it as a
film.
In this film form, KNO$_3$ has been proposed as a promising material for
random-access memory devices.\cite{Scott1987PRB}

\begin{figure}
\includegraphics[width=3.4in]{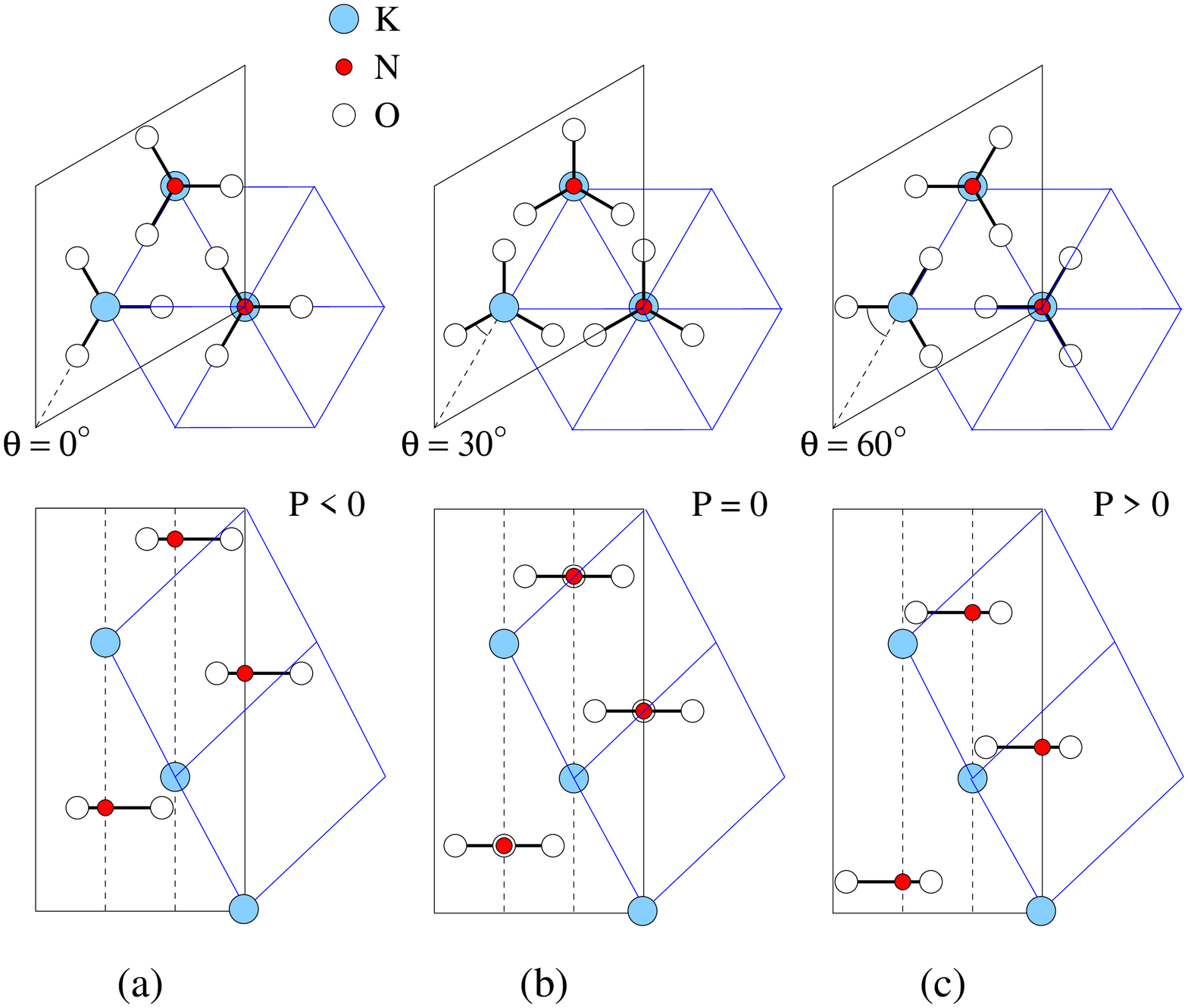}
\includegraphics[width=3.4in]{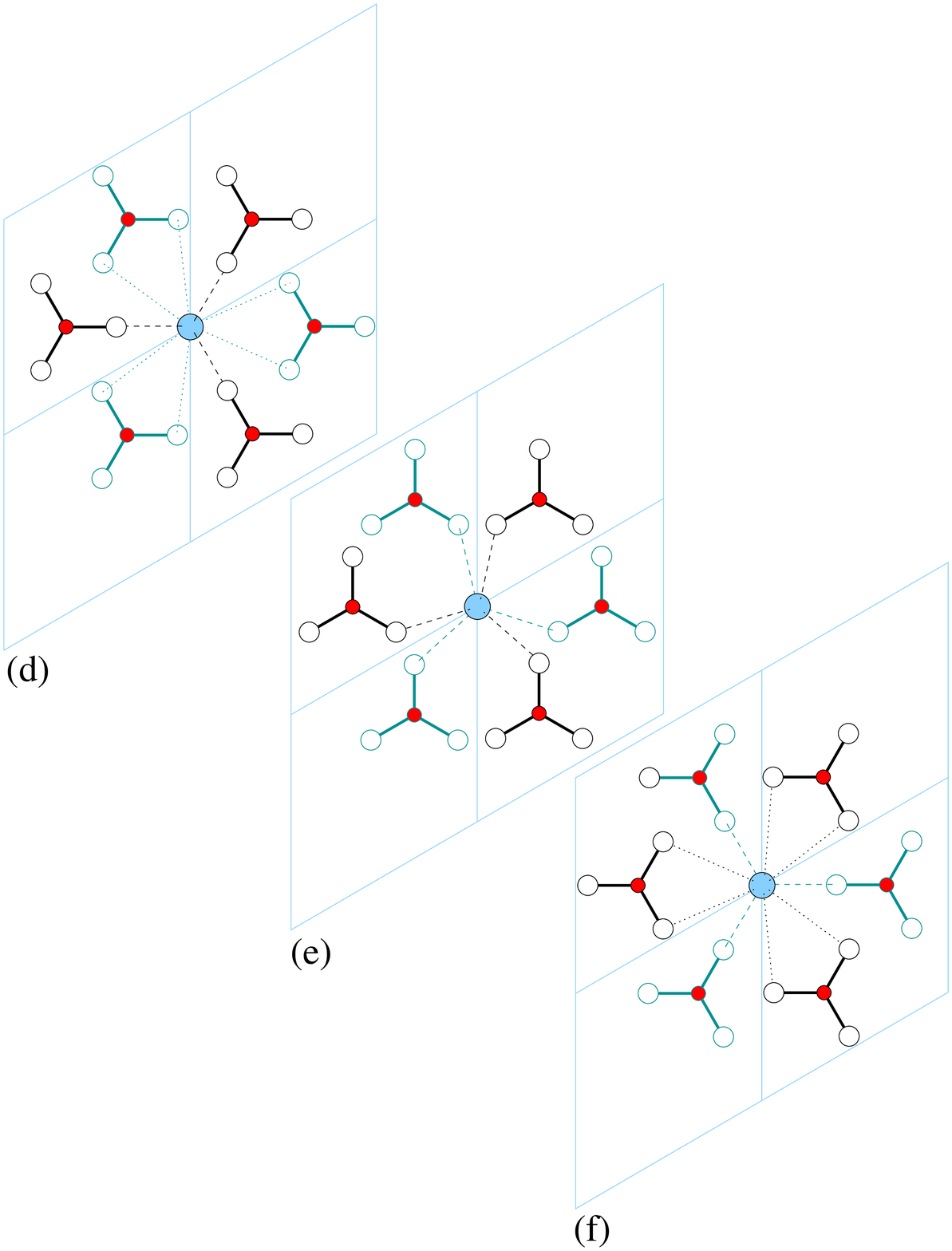}
\caption{(Color online) Top and side views of the fifteen-atom
conventional hexagonal cell (outlined in black) for ferroelectric KNO$_3$ in
(a) the ground state with negative polarization, (b) a metastable paraelectric
state, and (c) the ground state with positive polarization.
Blue lines indicate the equivalent five-atom rhombohedral cells.
Panels (d)-(f) show top views of the same structures as in (a)-(c),
clarifying the near-neighbor environment of a K atom. Neighbors in the
plane above and below are indicated with black and blue lines
respectively; first K-O neighbors are drawn as dashed lines, while
second K-O neighbors are designated by dotted lines.}
\label{fig_structures}
\end{figure}

The unit cell of phase III KNO$_3$ is a five-atom rhombohedral cell with
{\em R3m} symmetry as depicted in Fig.~\ref{fig_structures}(a).
This structure can be viewed as resulting from the stacking, along the
rhombohedral axis, of planes consisting alternately of K atoms and NO$_3$
groups.
However, it is found that a K plane is not equidistant from the 
NO$_3$ plane above it and the one below it.
Instead, the system finds it favorable to adopt a polar structure in which 
the distance between a given K atom and its six O second-neighbors in the
NO$_3$ plane below it is reduced at the expense of increasing the distance
between this same K atom and its three O first-neighbors in the NO$_3$ plane
above it [Fig.~\ref{fig_structures}(d)].
Experimentally, the optimum stacking occurs when every K plane is roughly three 
times closer to the NO$_3$ plane below it than to the one
above.\cite{Nimmo1976AC}
Of course, there must also exist a polarization-reversed structure
having the same energy but with the K plane closer to the NO$_3$
plane above than to the one below.  Moreover, the polarization reversal
has to be accompanied by a NO$_3$
rotation of 60$^\circ$ in order to preserve the symmetry.  The
resulting structure is shown in Figs.~\ref{fig_structures}(c) and (f).
A paraelectric configuration that is intermediate between these two
ferroelectric ones, having $R32$ symmetry, is shown in
Fig.~\ref{fig_structures}(b) and (e);
in this case the K planes are equidistant from the NO$_3$ planes above and
below, and each K atom has six equidistant O first neighbors.
This structure is not found to be stable in nature, but corresponds to
the saddle point on the energy surface connecting the degenerate
minima.
If we take the nitrate orientation in the case of Fig.~\ref{fig_structures}(a)
to correspond to an angle of $\theta = 0^\circ$, then in the case of
Fig.~\ref{fig_structures}(b) we have $\theta = 30^\circ$, and in the case
of Fig.~\ref{fig_structures}(c) we have $\theta = 60^\circ$.

We presented preliminary results for ferroelectric KNO$_3$ as an
illustrative example in a previous study in which we developed a
formalism for carrying out first-principles calculations
at constant polarization \cite{Dieguez2006PRL}.  We
obtained curves of energy $E$ versus polarization $P$ for polarizations
parallel to the rhombohedral axis of KNO$_3$
using not only the
full first-principles method, but also an earlier approximate
approach introduced by Sai, Rabe, and Vanderbilt (SRV).\cite{Sai2002PRB}
(The latter includes only the
lattice-displacement responses to the electric field, while the method
of Ref.~\onlinecite{Dieguez2006PRL} properly includes also the electronic
contributions and is therefore exact.)
We found that using the SRV method led to an $E(P)$ curve for
KNO$_3$ that was smooth, with the orientation of the NO$_3$ group
rotating continuously from $\theta = 0^\circ$ to
$\theta = 60^\circ$.  However, using the exact approach, we found
a quite different $E(P)$ curve having a cusp at zero polarization,
reflecting the presence of a discontinuity in $\theta(P)$ at $P=0$.
A brief discussion of this was given in Ref.~\onlinecite{Dieguez2006PRL},
where these features were explained through the introduction of a
simple model in which the energy of the system was expanded to
low order in the relevant degrees of fredom (the
polarization $P$ and the nitrate group rotation angle $\theta$).

The purpose of the present theoretical study is to investigate
the structural and dielectric properties of phase III of
KNO$_3$ in greater depth, and to provide a framework for understanding
this and related systems.
To do this, we combine extended first-principles calculations with
a detailed analysis of the analytical model to be presented shortly.
The picture that emerges is that of a system with a much richer variety of
possible behaviors than can be found in conventional ferroelectrics.
Some of these behaviors are realized for KNO$_3$ under external pressure,
while others correspond to conditions that we have not reached in our
first-principles calculations, but that could exist for related
materials.

The rest of this article is organized as follows. 
In Sect.~\ref{sect_model} we analyze the analytical model presented in 
Ref.~\onlinecite{Dieguez2006PRL}.
Section \ref{sect_abinitio} contains first-principles results for the 
ferroelectric phase of KNO$_3$ under pressure.
Finally, we summarize our results in Sect.~\ref{sect_summary}.

\section{Analytical model}
\label{sect_model}

In the model introduced in Ref.~\onlinecite{Dieguez2006PRL}, the energy
may be written as
\begin{eqnarray}
E(P,\theta) &=& E_0 + E^* \cos 6 \theta
\nonumber\\&&\quad
              + A \cos 12 \theta + B P \cos 3 \theta + C P^2
\label{eq_model_us}
\end{eqnarray}
where $E^*$ serves as an energy scale for the definition of a
dimensionless energy $\E=(E-E_0)/E^*$.  We also define a polarization
scale $P^*=\sqrt{E^*/C}$ and a dimensionless polarization $\P=P/P^*$,
and rewrite Eq.~(\ref{eq_model_us}) in rescaled form as
\begin{equation}
\E(\P,\theta) = \cos 6 \theta + \alpha \cos 12 \theta
              + \beta \P \cos 3 \theta + \P^2
\label{eq_model}
\end{equation}
where
\begin{equation}
\alpha = \frac{A}{E^*}
\end{equation}
and 
\begin{equation}
\beta = \frac{B}{\sqrt{CE^*}}
\end{equation}
are two dimensionless parameters determining the behavior of the
model.

\subsection{Phase diagram}

\begin{figure}
\includegraphics[width=2.8in]{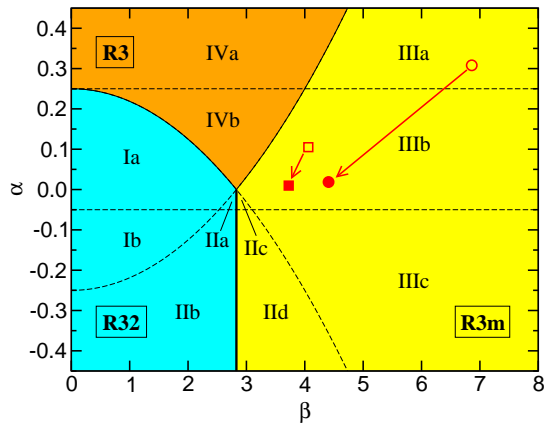}
\caption{(Color online) Phase diagram of the model given by
Eq.~(\ref{eq_model}) including
a paraelectric phase ($R32$) and two ferroelectric phases ($R3m$ and $R3$).
Each thick (thin) continuous line represents a first-order (second-order) phase 
transition.
The broken lines enclose areas of the diagram mentioned in the text.
The empty circle (square) corresponds to a fit of the exact
(SRV) $E(P)$ fixed-cell first-principles calculations for KNO$_3$
to Eqs.~(\ref{eq_model_us}-\ref{eq_model});
the filled symbols and the arrows indicate how the
situation changes when stress in the cell is relaxed (zero pressure).}
\label{fig_phasediag}
\end{figure}

We begin by analyzing in detail the possible behaviors that the model
system of Eq.~(\ref{eq_model}) can take as $\alpha$ and $\beta$ are
varied, thereby building up the phase diagram of Fig.~\ref{fig_phasediag}.
We will assume that $\beta > 0$, since the $\beta < 0$ case is equivalent
to changing $\P$ into $-\P$.
We identify the local minima of $E(\P,\theta)$ on the basis of its first
and second derivatives, arriving at the following classification.
First, there is a minimum at ($\P=0$, $\theta=30^{\circ}$) iff
$\alpha \leq (8-\beta^2)/32$ (areas I and II in Fig.~\ref{fig_phasediag}).
Second, there are two equivalent minima at ($\P=-\beta/2$, $\theta=0^{\circ}$)
and ($\P=+\beta/2$, $\theta=60^{\circ}$) iff $\alpha \geq (\beta^2-8)/32$
(areas II and III in Fig.~\ref{fig_phasediag}).
And third, there are two equivalent minima with rotation angles satisfying
$\cos 6\theta = (\beta^2 - 8) / (32 \alpha)$ and
$\P = - (\beta/2) \cos 3\theta$ iff $\alpha \geq (\beta^2-8)/32$ and
$\alpha \geq (8-\beta^2)/32$ (area IV in Fig.~\ref{fig_phasediag}).
Note that local minima of the first and second kind coexist in
area II, and we subdivide this according to whether the the global minimum
is of the first kind (for $\beta < 2\sqrt{2}$) or of the second
kind (for $\beta > 2\sqrt{2}$).  One is thus led to the phase
diagram depicted in Fig.~\ref{fig_phasediag}, showing three
ground-state phases: a paraelectric $R32$ phase (area I and the left part of
area II),
a ferroelectric $R3$ phase (area IV), and a ferroelectric
$R3m$ phase (the right part of area II and area III).
The transition from phase $R32$ to $R3m$ is of first order, while
the other transitions are of second order.

We establish a further subclassification of the areas in the diagram by
considering whether or not the corresponding $\E(\P)$ curves have continuous
derivatives (or equivalently, whether or not $\theta(\P)$ remains
continuous).  It can be shown that this depends only on $\alpha$.
For $\alpha > 1/4$, there is a discontinuity in $\theta(\P)$ at $\P = 0$,
with the angle jumping from $30^\circ-\theta_0$ to $30^\circ+\theta_0$,
where $0<\theta_0<30^\circ$.  For $-1/20<\alpha<1/4$, $\theta(\P)$ is
continuous and $\E(\P)$ is smooth.  Finally,
for $\alpha < -1/20$ there is a discontinuity in $\theta(\P)$ at
$\P=-\P'$ with $\theta$ jumping from 0 to
$\theta'$, and another equivalent discontinuity at a
positive $\P=\P'$ with $\theta$ jumping from 60$^\circ$ to
$60^\circ - \theta'$, where $\P'>0$ and a $0<\theta'<30^\circ$.
The areas that arise in this way have been labeled by appending a letter
(a, b, c, or d) to the Roman numeral used in the main classification.

\subsection{Triple-well potentials}

\begin{figure}
\includegraphics[width=2.8in]{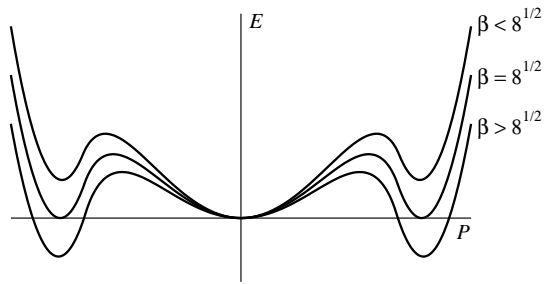}
\caption{Energy versus polarization curves in area IIa (top curve),
area IIc (bottom curve), and on the boundary separating them
(middle curve).  The value of $\alpha$ is the same in all three cases; the
system goes from paraelectric to ferroelectric with increasing $\beta$.}
\label{fig_triplewell}
\end{figure}

A very interesting area of the phase diagram of Fig.~\ref{fig_phasediag}
is area II, in which a minimum at ($\P=0$, $\theta=30^{\circ}$) coexists
with
the pair of equivalent minima at $\theta=0^{\circ}$ and $\theta=60^{\circ}$.
It can be shown that if $\beta < 2\sqrt{2}$ then the paraelectric minima has
the lowest energy, while if $\beta > 2\sqrt{2}$ the ferroelectric minima
are energetically favoured.
Curves of $\E(\P)$ for both cases, as well as for the $\beta = 2\sqrt{2}$ case,
are drawn in Fig.~\ref{fig_triplewell}.
They correspond to areas IIa and IIc, in which the derivative for each of
them
is continuous, and illustrate clearly the first-order character
of the transition from the $R32$ phase to the $R3m$ phase. (Analogous curves
can be found at areas IIb and IId, but in those cases there are two points
at which the first-derivative of $\E(\P)$ is not continuous.)
These curves are similar to the ones that describe the free energy
of a ferroelectric material at temperatures in the vicinity of a
first-order paraelectric-ferroelectric transition (see, e.g.,
Ref.~\onlinecite{Lines1977book}).  However, we are not aware of any previous
instance in which a first-principles mapping of energy versus polarization
has revealed this kind of triple-well behavior at zero temperature.

\subsection{Fitting previous first-principles results}

\begin{figure}
\includegraphics[width=2.6in]{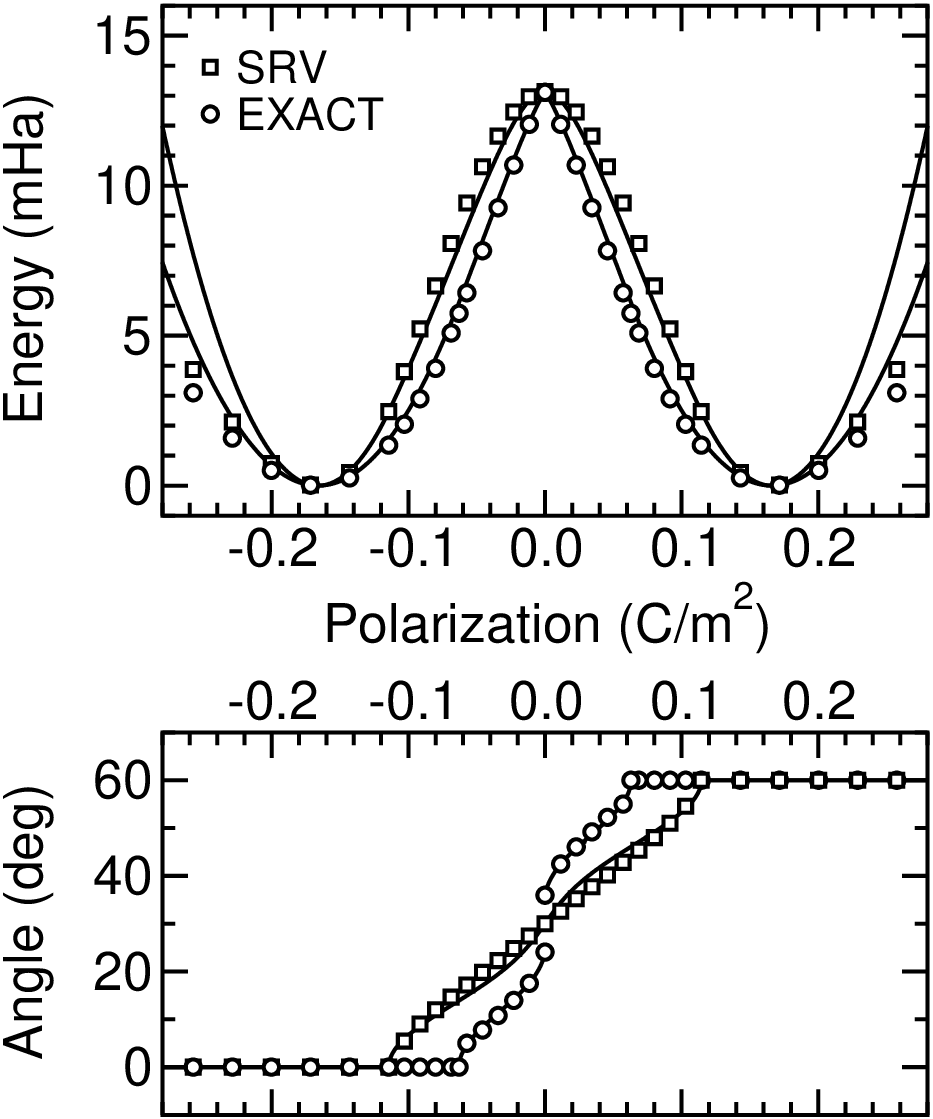}
\caption{Energy versus polarization, and NO$_3$ rotation angle versus
polarization, for the ferroelectric phase of KNO$_3$ without 
strain relaxation.
The squares and circles represent SRV and exact first-principles calculations
respectively (from Ref.~\onlinecite{Dieguez2006PRL}),
while the curves have been fitted to these data using
Eqs.~(\ref{eq_model_us}-\ref{eq_model}) as described in the text.}
\label{fig_curves_norelaxed}
\end{figure}

The SRV results in Ref.~\onlinecite{Dieguez2006PRL} provide an example of a system
showing a behavior that belongs in area IIIb, while the exact results case
falls into area IIIa.  Fitting the first-principles results to
Eq.~(\ref{eq_model_us}) and using the rescaling of Eq.~(\ref{eq_model}),
we find $\alpha = 0.105$ and
$\beta = 4.06$ for the SRV case, while instead $\alpha = 0.308$ and
$\beta = 6.86$ for the exact case.
These points have been indicated in Fig.~\ref{fig_phasediag} with an empty
square and circle, respectively.
The model curves that result when these values are used, together with the
points obtained from first-principles calculations, are plotted in
Fig.~\ref{fig_curves_norelaxed}.
We can see that in the region between the two minima the fit is excellent,
indicating that our relatively simple model captures the important $E(P)$
physics of the system.
As the absolute magnitude of $P$ grows too large, more powers
of $P$ are needed in the model to improve the fit.
We have found that including terms up to fourth order
works very well even for large values of $P$.

\section{First-principles results}
\label{sect_abinitio}

\subsection{Cell vectors relaxation}

The calculations in Ref.~\onlinecite{Dieguez2006PRL} were carried out with the
lattice vectors fixed to be those of the ground state, even when the
polarization is varied away from its spontaneous value.
It is also possible to relax the cell vectors;\cite{Wu2007inprep}
we have now done that here, arriving at results valid
for the more realistic situation of stress-free (i.e., zero-pressure)
boundary conditions.  As in our previous work,\cite{Dieguez2006PRL} we use
density-functional theory \cite{Hohenberg1964PR} as implemented in the ABINIT
code, \cite{abinit} the local-density approximation 
(LDA), \cite{Ceperley1980PRL} a plane-wave cutoff of 30 Ha, a
reciprocal space grid with 6 inequivalent points, and
Troullier-Martins \cite{Troullier1991PRB} pseudopotentials.\cite{pseudosKNO3}

\begin{figure}[t]
\includegraphics[width=2.6in]{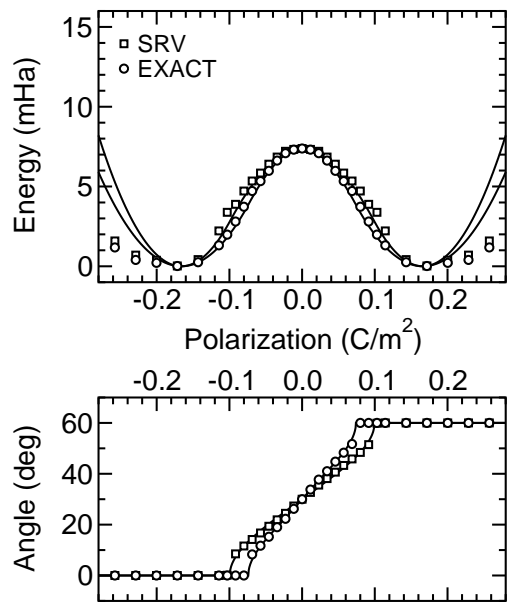}
\caption{Energy versus polarization, and NO$_3$ rotation angle versus
polarization, for the ferroelectric phase of KNO$_3$ 
with strain relaxation.
The squares and circles represent SRV and exact first-principles calculations
respectively,
while the curves have been fitted to these data using
Eqs.~(\ref{eq_model_us}-\ref{eq_model}) as described in the text.}
\label{fig_curves_relaxed}
\end{figure}

The fitting parameters $\alpha$ and $\beta$ resulting from these new
calculations are shown in Fig.~\ref{fig_phasediag} as filled
symbols, while the corresponding $E(P)$ and $\theta(P)$ curves
are shown in Fig.~\ref{fig_curves_relaxed}.
These values are compared with 
the ones for the non-relaxed case in Table \ref{tab_alphabeta}.
The SRV results are similar to the ones in the non-relaxed case, apart from 
a reduction in the height of the double-well barrier that might have been
expected in view of the additional degrees of freedom that are relaxed
in the free-stress case.  However, the exact results have changed
qualitatively.  The cusp at zero polarization and the corresponding
discontinuity in the NO$_3$ rotation angle have now
disappeared, and the exact curves are quite similar to the SRV ones.
We have also found that as the polarization crosses through zero,
the cell volume expands appreciably.  Specifically, the
rhombohedral lattice parameter goes from 7.62\,a.u.\ in
the ferroelectric ground state to 8.00\,a.u.\ in the paralectric case (a 5\% 
increase), with the rhombohedral angle going from 78.0$^\circ$ to 76.8$^\circ$.
This increase in lattice volume occurs so that the N-O bond length
\begin{table}[b]
\caption{Values of the model parameters $\alpha$ and $\beta$ depending on 
whether the cell vectors are relaxed or not, and whether we use the SRV
method or the exact one.}
\begin{center}
\begin{tabular}{lcccc}
\hline \hline
                      & \multicolumn{2}{c}{SRV} & \multicolumn{2}{c}{Exact} \\
                      &  $\alpha$  &  $\beta$   &  $\alpha$  &  $\beta$     \\
\hline 
Fixed cell vectors    &   0.105     &  4.06      &  0.308     &  6.86       \\
Relaxed cell vectors  &   0.010     &  3.72      &  0.019     &  4.41       \\
\hline \hline
\end{tabular}
\label{tab_alphabeta}
\end{center}
\end{table}
can keep similar values in the ferroelectric $R3m$ ground state 
(2.353\,a.u.) and in the less closely-packed paraelectric $R32$ state
(2.351\,a.u.).  The reason for the appearance of the cusp at $P$=0 in the
case of fixed lattice constants of 7.62\,a.u.\ can be traced to the
fact that the N-O bond distance becomes uncomfortably short at
2.337\,a.u.\ if the $R32$ symmetry is enforced.  Instead, the system
prefers to lower its symmetry to $R3$ via a rotation of the nitrate
groups (from $\theta=30^\circ$ to $36.1^\circ$ or $23.9^\circ$,
depending on whether $P\rightarrow0^+$ or $0^-$),
with the N-O distance attaining a slightly larger length of 2.338\,a.u. 

\begin{figure}
\includegraphics[width=3.2in]{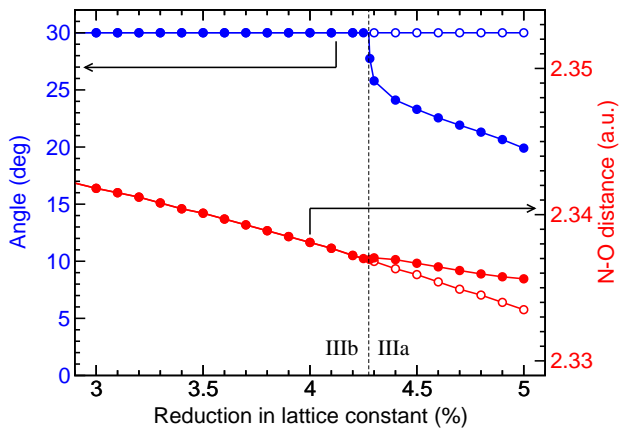}
\caption{(Color online) Zero-polarization first-principles calculations of the value of the 
NO$_3$ rotation angle and of the N-O distance in KNO$_3$ as functions of 
the reduction in the lattice parameter.
The empty symbols correspond to 
calculations in which the $R32$ structure is imposed, while the the filled
symbols correspond to calculations in which this restriction is removed.
All calculations were done using the exact method of 
Ref.~\onlinecite{Dieguez2006PRL}.}
\label{fig_pressure}
\end{figure}

\subsection{Calculations under pressure}

The description provided in the previous paragraph seems to indicate that if
the ferroelectric phase of KNO$_3$ were compressed, we should expect to see 
the $E(P)$ cusp and $\theta(P)$ discontinuity at zero polarization reappearing 
at some value of applied pressure.
To test this hypothesis, we have performed zero-polarization calculations 
for lattice parameters smaller than $8.00$\,a.u.\ while keeping the
rhombohedral angle fixed at 76.8$^\circ$.
This is computationally simpler than doing calculations at constant
pressure, and we have confirmed that the anisotropy in the stress tensor
obtained in this way remains quite small ($\sim$3 GPa at most),
so that the rhombohedral angle would be expected to change very little if
the rhombohedral angle were relaxed.  The results, shown in
Fig.~\ref{fig_pressure}, indicate that when the lattice parameter is reduced
by around 4.3\%, a cusp reappears in the $E(P)$ curve and the $R32$ structure
becomes unstable to the $R3$ structure ($\theta\ne30^{\circ}$.) at $P=0$.
The figure also shows that the N-O bonds are not so tightly
compressed as they would be at $\theta=30^{\circ}$.
Therefore, we recover the IIIa and IIIb behaviors described in 
Ref.~\onlinecite{Dieguez2006PRL}, but this time it is the pressure applied
to the material that drives the transition between them.

\section{Summary}
\label{sect_summary}

To summarize, we report a theoretical study of the properties of ferroelectric
phase III of KNO$_3$.
We have carried out first-principles calculations that show that
the polarization reversal mechanism in this material is accompanied by a
rotation of 60$^\circ$ in the orientation of the nitrate groups.
When the pressure exerted on the system is high enough, the corresponding
double-well potential shows a distinctive cusp at zero polarization,
and the mentioned rotation is discontinuous.
These features are very well reproduced by a simple model in which the energy
is expanded to low order in the polarization and the rotation angle of the
nitrate groups.
The phase diagram that arises from this model contains regions that can be
realized by varying the pressure on ferroelectric KNO$_3$, as well as
other regions that might turn out to be relevant for related materials
in the future.  One particularly interesting region of the phase diagram
describes a system having a triple-well structure as a function of
polarization, a feature not yet encountered in the context of
first-principles studies of ferroelectric energy landscapes.

\begin{acknowledgments}
The authors thank J.~Scott for suggesting the study of KNO$_3$.
This work was supported by ONR Grant N0014-05-1-0054.
\end{acknowledgments}



\end{document}